\documentclass[aps,twocolumn]{revtex4}
\usepackage{graphicx,color}

\begin{document}
\title{Laboratory experiments on the generation of internal tidal beams over steep slopes}

\author{Louis Gostiaux and Thierry Dauxois} \date{\today}
\affiliation{Laboratoire
  de Physique, ENS Lyon, CNRS, 46
  All\'{e}e d'Italie, 69364 Lyon c\'{e}dex 07, France\\
  \email{Louis.Gostiaux@ens-lyon.fr,Thierry.Dauxois@ens-lyon.fr}}
\bibliographystyle{plain}

\begin{abstract}
We designed a simple laboratory experiment to study internal tides
generation. We consider a steep continental shelf, for which the
internal tide is shown to be emitted from the critical point, which
is clearly amphidromic. We also discuss the dependence of the width
of the emitted beam on the local curvature of topography and on
viscosity. Finally we derive the form of the resulting internal
tidal beam by drawing an analogy with an oscillating cylinder in a
static fluid.
Hydrodynamic waves.
\end{abstract}
\maketitle

Over the past two decades it has become apparent that substantial
internal tides can be generated by tidal currents over ridges and
other rough topography of the ocean floor.  This problem is of
paramount importance since baroclinic tides, generated by barotropic
currents over ocean ridges and seamounts, are an important source of
the ocean interior mixing. Indeed, recent observations suggest that
mixing in the abyssal oceans is rather weak, except in localized
regions near rough topography. It explains the discrepancy between
the observed intensity of mixing in the interior of the oceans and
is required to satisfy models of ocean circulation~\cite{BMW00}.
This suggests that these topographic effects must be incorporated
into realistic climate and circulation models.

\begin{figure}[!ht]
\centering
\includegraphics[width=0.45\textwidth]{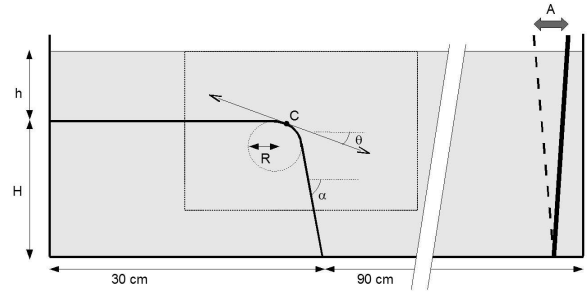}
\caption{Sketch of the experimental set-up. A PVC
plate is oscillating between the solid and dashed oblique lines
indicated on the right, creating the tide.  The continental shelf
consists of two planar PVC plates connected by a quarter of cylinder
of radius $R=3.3$ cm. The upper plate is nearly horizontal, while
the downgoing one makes an angle $\alpha=78^\circ$ with the
horizontal. The aspect ratio $h/H$ is 2/3, for a total water height
$h+H=19$ cm. The critical point C indicates the location where the
internal tide generation is supposed to occur, i.e. where the
internal beam inclined at an angle $\theta=23^\circ$ with the
horizontal is tangent to the topography.  The dotted rectangle
indicates the  zone captured by the camera and presented in
Fig.~\ref{fig:harmonic1}a.} \label{fig:experimentalsetup}
\end{figure}

This question has been studied theoretically and numerically, while
related oceanic observations have been reported. Baines
showed~\cite{B82} first that points where the topographic slope
coincides with the angle of propagation might be an effective
generator of internal wave. Such a point is referred to as the
critical point (see Fig.~\ref{fig:experimentalsetup}).
Internal beams emanating from the continental slope have been
observed recently~\cite{PN91,NP92,JMP02,GLM04,ASN06}.  On the
theoretical side, let us mention several interesting
descriptions~\cite{SSGP03,PLY06} of abrupt discontinuity cases such
as a topographic step, a knife-edge ridge (zero width ridge as a
simple model for tall ridges such as the Hawaiian one), a
tent-shaped ridges, or smooth ridges (gaussian, polynomial).
Finally, recent numerical reports have also provided some new
insight about this mechanism~\cite{GLM04,GSBA06}.

All these works converge to conclude that beams of internal tides
energy arise due to the interplay between oscillating currents and
bathymetric features.  Of particular interest is the location where
the slope is critical, i.e. where the direction of the wave beam is
tangent to the slope.  In the remaining of the paper, we
experimentally study the location of emission and the form of the
internal beam which is emitted. We also explain how this form can be
predicted by drawing an analogy between the emission of an internal
tide from a static topography, and the emission of internal waves by
an oscillating cylinder in a static fluid. The agreement is very
good, even quantitatively.


The experiment was performed in a rectangular-shaped tank (length
120~cm, height 40~cm and width 10~cm) as sketched in
Fig.~\ref{fig:experimentalsetup}.  A curved thin (0.5~cm) PVC plate,
introduced before filling, played the role of the continental shelf.
The experiment was performed with a linear stratification, pure
water at the free surface while highly salted water at the bottom,
resulting in a constant Brunt-Vais\"al\"a buoyancy
frequency~$N\simeq 0.81$~rad/s. In the absence of rotation, the
dispersion relation of internal gravity waves
 reads $\omega=N\sin\theta$, where $\theta$ is the angle of the beam with respect to the horizontal.

The tidal current was generated by a second PVC plate localized far
from the continental slope, i.e. in the region corresponding to the
abyssal plane or to the open ocean.  This plate was fixed at its
bottom end, with its top end oscillating horizontally. The
oscillation $A\sin(\omega t)$ was monitored by a position control
motor of a printer apparatus driven by a sinusoidal electronic
signal delivered by a Kepco amplifier. The phase origin at $t=0$
corresponds to a vertical position of the paddle and the movement is
initiated leftwards towards the topography. Several forcing tidal
frequencies and amplitudes were tried out. The tidal frequency
$\omega$ was satisfying the condition $0<\omega<N$, so that internal
tides may be freely radiating. Results and figures discussed
throughout this paper correspond to $\omega/(2\pi)=0.050$~Hz for
which $\theta=23^\circ$, and $A=0.5$~cm.

The density gradient field within the tank was finally obtained with the standard synthetic
Schlieren technique~\cite{DHS00} by acquiring successive side views
with a CMOS AVT Marlin F131B Camera. A sheet of paper with points
randomly scattered was located 150~cm behind the tank. Successive images
($1280\times300$ pixels) of those points obtained with the camera located
280~cm in front of the tank were adequately treated by a homemade
software.  A correlation image velocimetry algorithm~\cite{FD00} was
applied between a reference image taken prior to the experiment and
the different snapshots. As we used $21\times21$ pixels correlation boxes
with a 75$\%$ overlap, the resolution for the two dimensional
density gradient field is $256\times60$.

\begin{figure}[!ht]
\flushleft{\bf a)}\\
\vskip -0.8 truecm \centering
\includegraphics[width=0.5\textwidth]{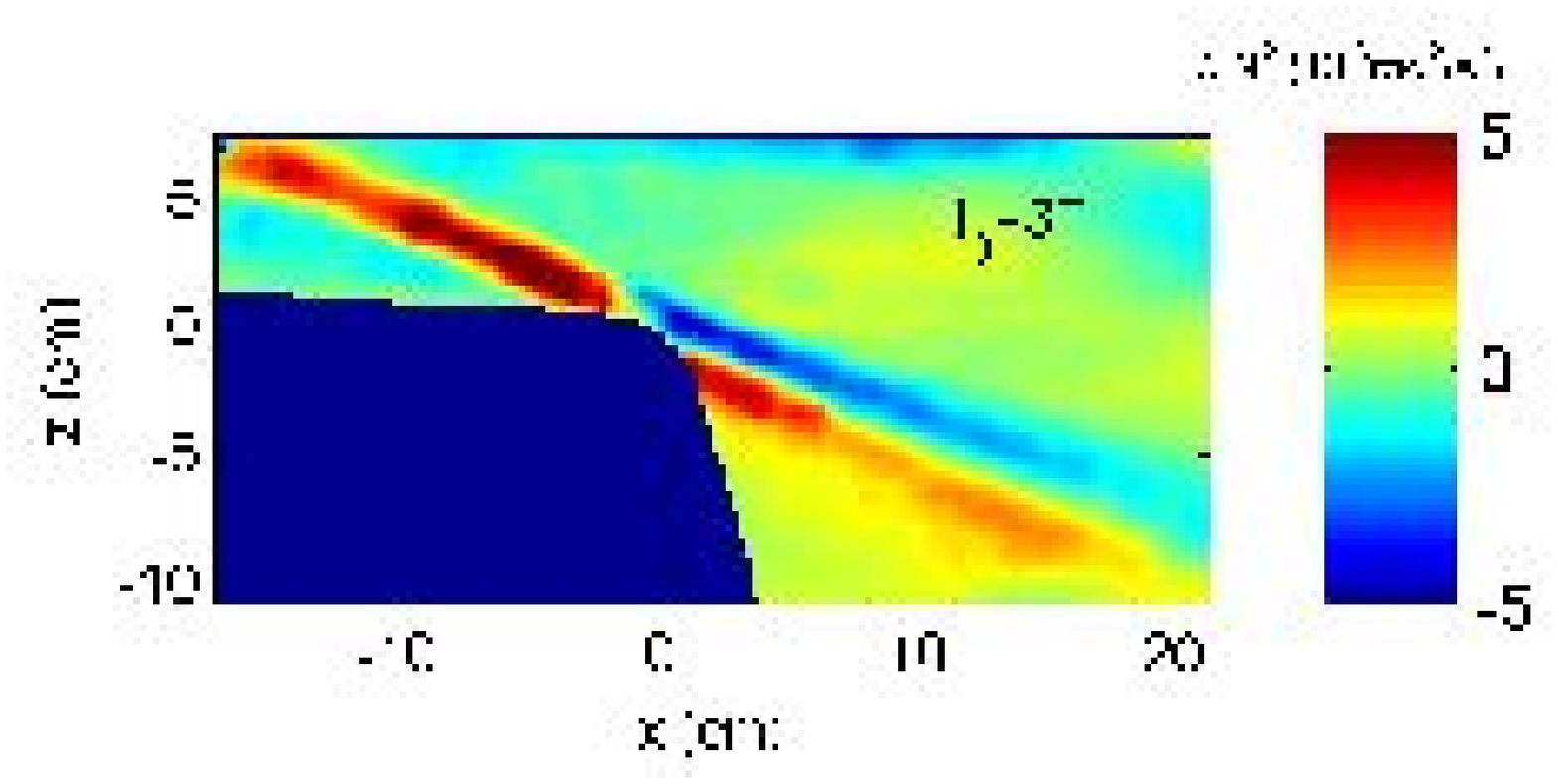}
\flushleft {\bf b)}\\
\vskip -0.8 truecm\centering
\includegraphics[width=0.5\textwidth]{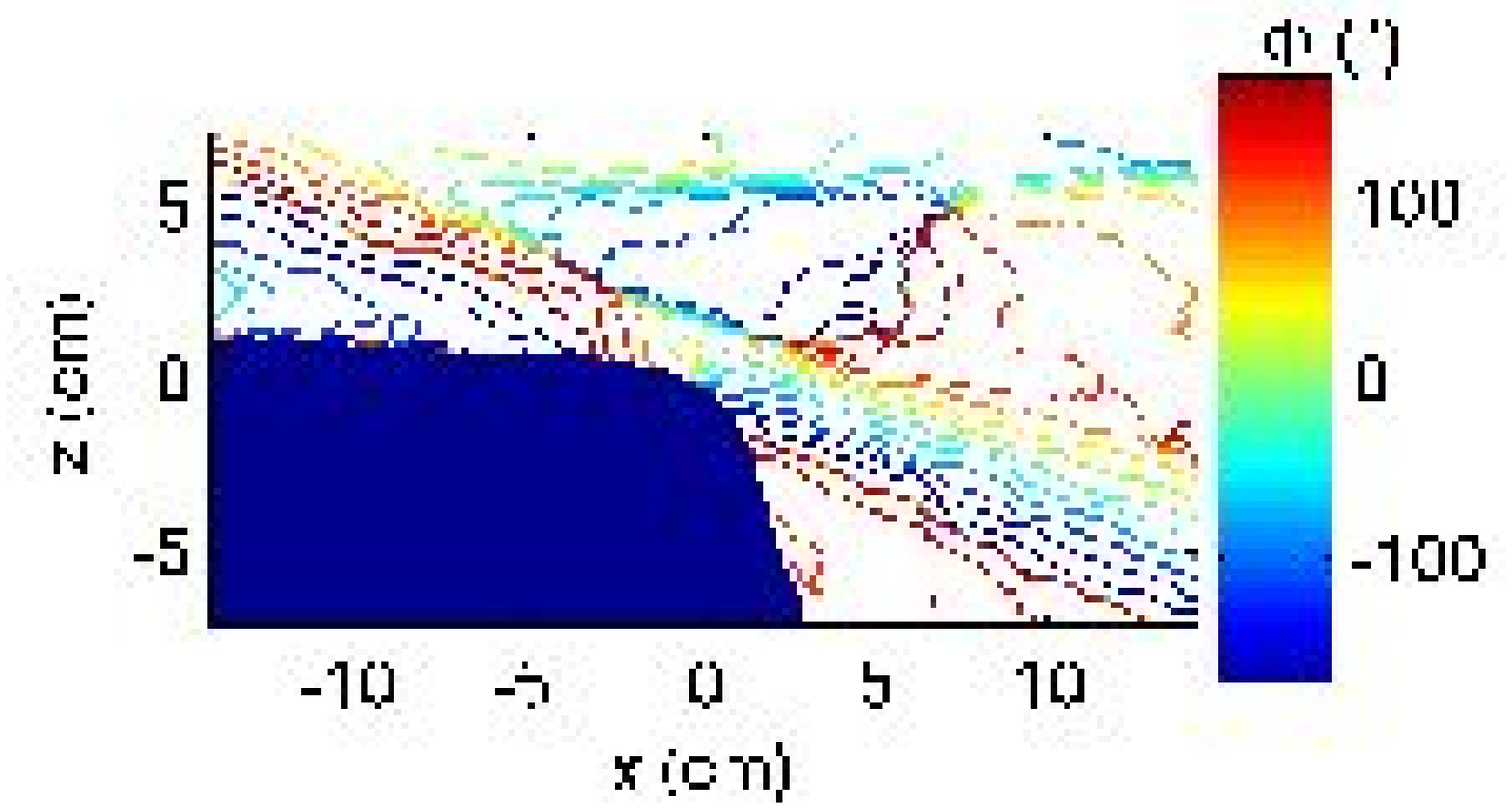}
\centering \caption{(Color online) Panel a presents the
two-dimensional vertical density gradient, while panel b shows the
phase. Both plots have been averaged over one tidal period of the
experiment around $t=3T$. Horizontal and vertical distances are in
cm. The generation point C appears to be an amphidromic point around
which the phase of the wave rotates.} \label{fig:harmonic1}
\end{figure}

Figure~\ref{fig:harmonic1}a show a snapshot of the vertical density
gradient.  To facilitate the visualization, a narrower domain than
in the experiments is shown.  It corresponds to the dotted rectangle
depicted in Fig.~\ref{fig:experimentalsetup}. To increase the
visibility, we filtered at the excitation frequency $\omega$ the
time series of the density gradient field over one experimental
tidal period (see Ref.~\cite{GDDSV06} for more details about this
method).


An internal tide is clearly seen to emanate from the upper part of
the slope  and radiate away from the shelf-break in both directions.
However, it is important to stress the absence of a third beam
radiated transversally to the topography. This contradicts Baines
analytical theory~\cite{B82} in which this third beam reflects on
the surface and is present in the general solution for the downward
propagating wave. The explanation presumably lie in the presence of
a singular point in Baines' case or in Saint Laurent et al numerical
simulation~\cite{SSGP03}. In numerical experiments with a smooth
slope and a typically shallow continental shelf~\cite{GLM04}, no
such beam was found and the presence of this third beam was already
ambiguous in previous experiments~\cite{BF85}.

The amplitude of the vertical density gradient is given in terms of
variations of the squared Brunt-V\"ais\"al\"a frequency. If one
considers the original value of $N^2=0.66$~rad$^2$/s$^2$, the
measured amplitude of $\Delta N^2=\pm0.005$~rad$^2$/s$^2$ for the
internal wave is a very small perturbation of the original
stratification. Using the mass conservation relation
\begin{equation}\label{eq:rho}
i\omega(\rho-\bar{\rho})=w\frac{\mbox{d} \bar{\rho}}{\mbox{d} z}\,,
\end{equation}
where $\rho$ is the perturbated density, $\bar{\rho}(z)$ the initial
density and $w$ the vertical density, we obtain typical vertical
velocities of order $\pm 0.11$~mm/s, corresponding to vertical
displacements of $\pm 0.15$~mm. This is one of the interests of the
synthetic Schlieren technique that allows to measure very weak
perturbations of the buoyancy field and thus to investigate weakly
nonlinear regimes. This vertical amplitude has to be compared to the
barotropic elevation of the water induced by the forcing. The paddle
oscillates with an $A=0.5$~cm amplitude, and the width of the free
water volume in the tank is $90$~cm for a height of $H=12$~cm. The
corresponding elevation of the water at the level of the slope is
thus  $\Delta H=0.7$~mm. The baroclinic component observed is thus
still a perturbation of the barotropic tide.



To the right of the generation point C, energy propagation is
downward while, to the left, it is upward. As a consequence of the
internal waves propagation law for which group and phase velocities
are orthogonal with opposite vertical components, phase propagation
is thus upward to the right and downward to the left. One can
conclude from this simple observation that the phase has to rotate
around C, which is therefore an amphidromic point. Our filtering
technique allows to evaluate the phase of the wave~\cite{GDDSV06},
which is plotted in Fig.~\ref{fig:harmonic1}b. One can clearly see
that the isophase lines converge on a single point previously
referred as the generation point C, around which the phase rotates
uniformally.

Whereas the location and the inclination of the internal tide are
well understood, the selection mechanism of the width of the beam
was not yet clearly identified by previous studies. Several length
scales can be considered in this problem. The first one corresponds
to the thickness of the oscillating boundary layer
$\delta=(\nu/\omega)^{1/2}$ where $\nu$ is the kinematic viscosity.
In the present case, $\delta=1.8$~mm. The second one is the local
radius of curvature of the continental shelf $R=3.3$~cm. At last,
the dimensions of the shelf itself ($h$, $H$...) that play a
role~\cite{B82} in the ``flat-bump" geometry for which
$\alpha<\theta$ happen to be irrelevant in the configuration of a
steep topography.

Our understanding is directly inspired from the generation of
internal waves by oscillating cylinders.  One can therefore try to
draw an analogy between the internal tide generation by a curved
static topography of a given radius of curvature $R$ and the
internal waves generation by an oscillating cylinder of the same
radius $R$ in a stratified fluid.

\begin{figure}[htb]
\flushleft {\bf a)}\vskip -0.8truecm  \centering
\includegraphics[width=0.5\textwidth]{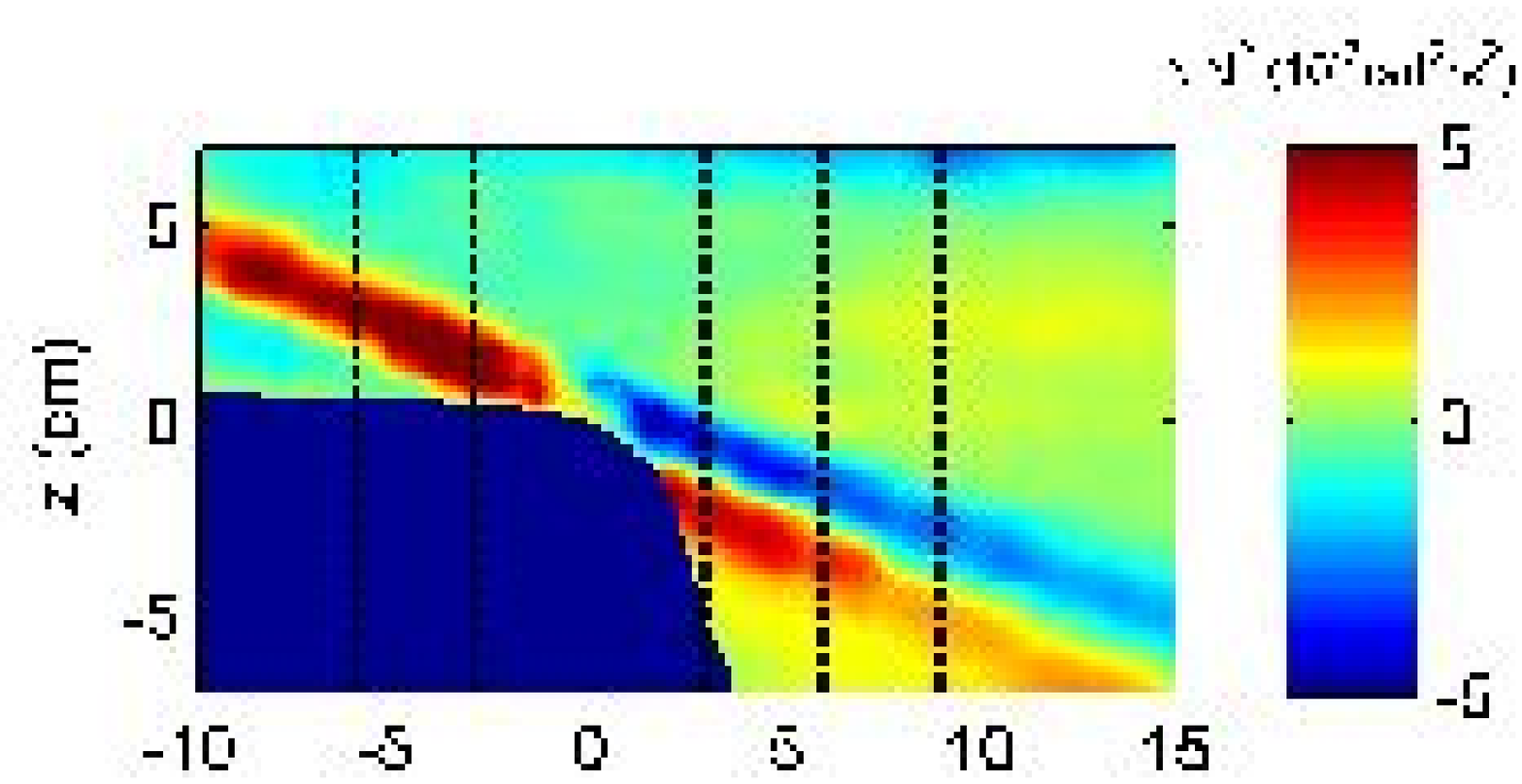}
\flushleft {\bf b)}\vskip -0.8truecm
 \centering
\includegraphics[width=0.5\textwidth]{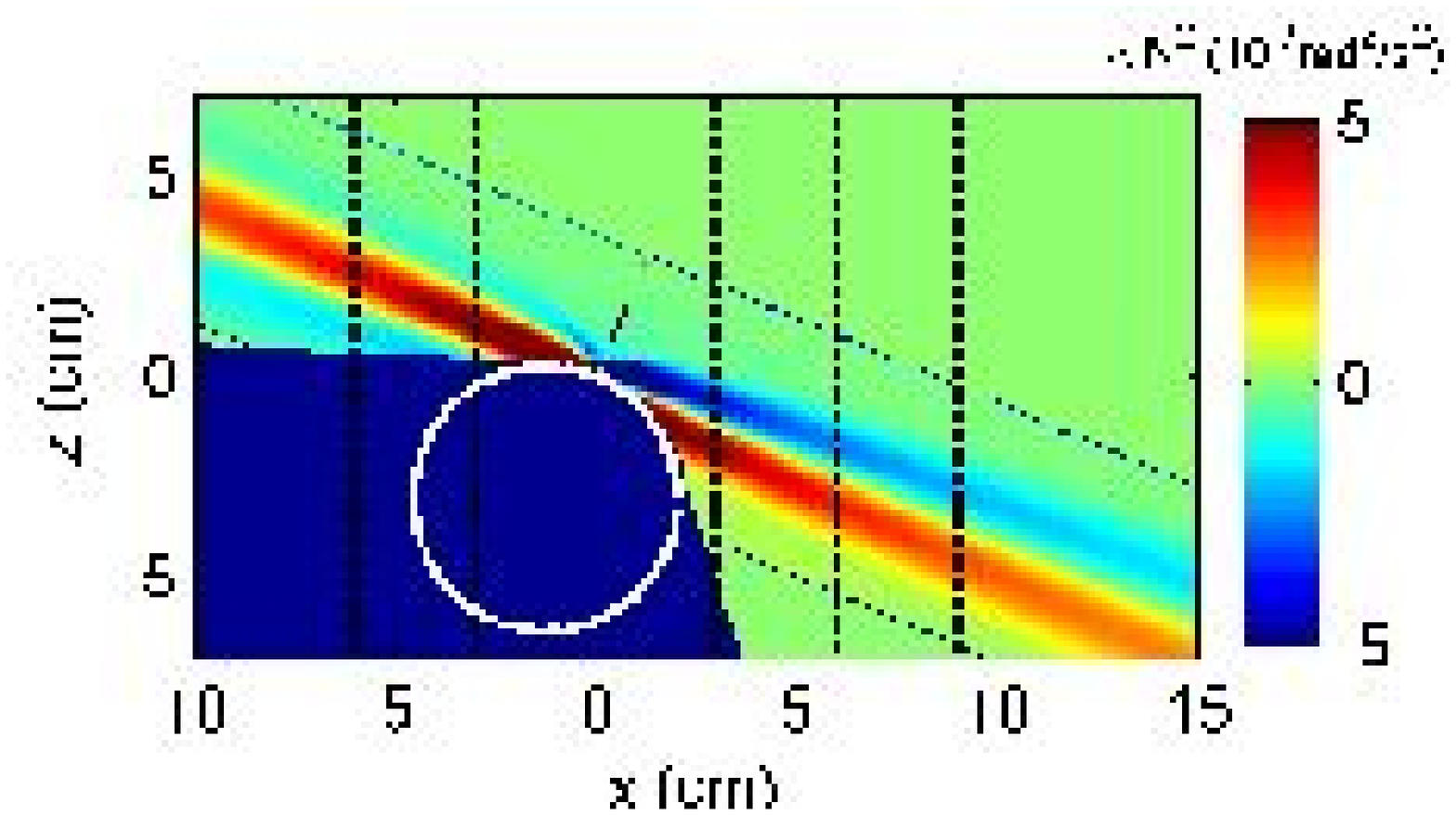}
\caption{(Color online) Comparison between experiment (a) and theory
(b) for the vertical density gradient field. Panel (a) has been
averaged over one tidal period around $3T$. We use the Hurley and
Keady analytical solution~\cite{HK06} for oscillating cylinders to
modelize the internal wave beam. We consider the tangential circle
at the shelf break as the generating cylinder section and compute
the analytical solution isolating the upper right emission point.
Vertical dashed lines correspond to the cross section plotted in
Fig.~\ref{fig:Keady2}.}\label{fig:Keady}
\end{figure}

In their seminal theoretical work~\cite{HK06}, Hurley and Keady
showed how to get the velocity profile of the beams generated by
such a cylinder.  First, they proposed  to consider a dimensionless
parameter $\lambda=\nu/(2R^2\omega\cot \theta)$ which is
proportional to the squared ratio of the oscillatory boundary layer
thickness $\delta$ that surrounds the cylinder to its radius~$R$.
The cylinder emits the four beams of the well known St Andrew's
configuration. Each beam is considered independently. Then, the
longitudinal velocity component~$v_s$  for each beam is shown to be
given by
\begin{equation} v_s(s,\sigma)=V_0\int_0^{+\infty}\!\!J_1(K)\exp(-\lambda s
K^3/R+i\sigma K)\, \mbox{d} K,
\end{equation}
where $J_1$ is the Bessel function of the first kind of order~1 and
$s$ (resp. $\sigma$) the longitudinal (resp. transversal) coordinate
along (resp. across) the beam. $V_0$ is the projection of the
oscillation velocity along the beam. In the domain
$\delta=(\nu/\omega)^{1/2}\ll R$, this function happens  to be
localized on the characteristics $\sigma=\pm R$ corresponding to the
lines of slope $\tan\theta$, tangential to the cylinder. In this
limit case, $2\times4=8$ beams are emitted from the four critical
points of the cylinder.

In our case, we only consider the upper characteristic corresponding
to the single critical point C of the problem. Hurley and Keady's
expression restricted to this single characteristic happens to fit
very well our experimental data. In the tidal experiment,
$\lambda\simeq6.3\ 10^{-4}$. We estimate the buoyancy perturbation
of the stratification by means of Eq.~(\ref{eq:rho}). By
appropriately taking into account the inclination of the beam, one
gets a final expression for the vertical density gradient that can
be compared to our experimental data.

Fig.~\ref{fig:Keady} compares the vertical density gradient field in
the experimental (a) and the theoretical (b) cases. The analytical
solution of Hurley and Keady is computed on an imaginary cylinder
tangential to the shelf break region. The amplitude of oscillation
of the cylinder corresponds to the amplitude of the tidal flow. The
agreement is qualitatively very good. The phase opposition between
the two parts of the beams is also well described by the model.

\begin{figure}[ht]
\includegraphics[width=0.45\textwidth]{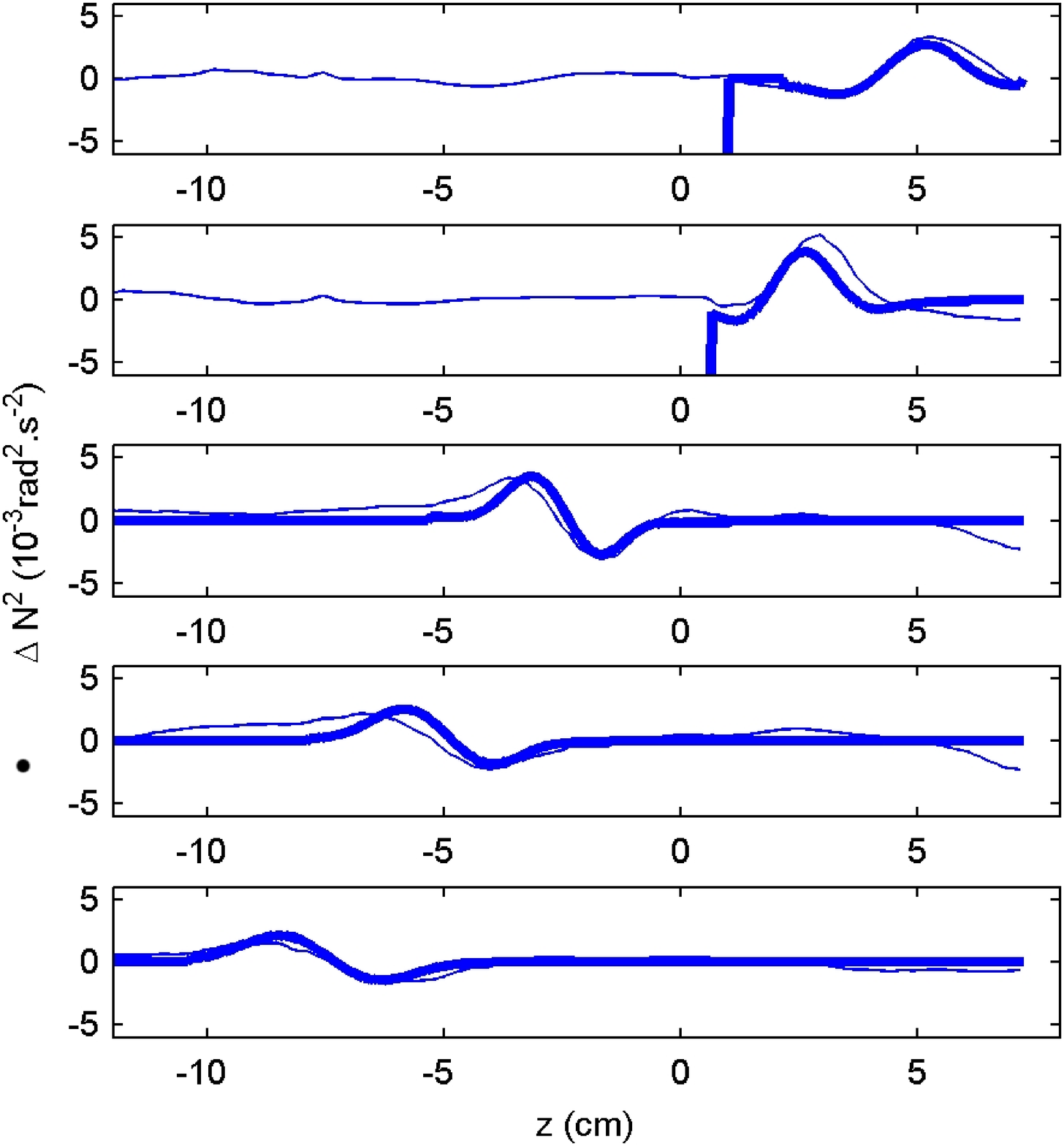}
\caption{Vertical cross sections of the measured vertical density
gradient (thin line) and its theoretical prediction using Hurley and
Keady's theory (bold line). From top to bottom, the cross sections
correspond to the abscisse $x=-6$, $-3$, 3, 6 and
9~cm.}\label{fig:Keady2}
\end{figure}

To get a more quantitative comparison between experiment and theory,
we plot on Fig.~\ref{fig:Keady2} five vertical cross sections
(indicated in Fig.~\ref{fig:Keady}) of the vertical density gradient
field at $x=-6$, $-3$, 6 and 9~cm relative to the generation point.
The quantitative agreement between the experimental and the
theoretical profiles is very good, which confirms that the structure
of the emitted beam only depends on the viscous boundary layer
thickness $\delta$ and on the local radius of curvature of the shelf
$R$.



First of all, this experiment proves that the internal tide
generation conditions are met when the radiated tidal beam is
aligned with the slope of the topography, corresponding to a
critical point of emission which happens to be amphidromic. We
clearly demonstrate the absence of any transverse emission at that
point that is supposed to occur in Baines analytical
study~\cite{B82}. It is however important to emphasize that this
disagreement might come from the absence of any sharp corner in the
shelf break, contrary to what has been considered in
Refs~\cite{B82,BF85,SSGP03}. Following this observation, we derive a
simple analytical model derived from the theoretical work of Hurley
and Keady~\cite{HK06} and show that we can quantitatively estimate
the density perturbation profile as a function of the viscosity and
the local radius of curvature of the topography.

Since this paper focuses on the internal tide generation, we dot not
consider the evolution of the internal tides as they propagate
further into the deep ocean.
 Surface reflections, in particular, are expected to
generate internal solitary waves by creating local disturbances at the
seasonal thermocline~\cite{NP92,G01}.

Moreover, the emitted internal wave of tidal period often breaks up
into internal waves of shorter period. Indeed, and very
interestingly, a recent numerical report has emphasized that
parametric subharmonic resonance may come into play in the rotating
case~\cite{GSBA06}. Experimental
work along those lines is in progress.

{\bf Acknowledgments}

We thank F. Petrelis and S. Llewellyn Smith for helpful discussions.
Comments to the manuscript by Denis Martinand are deeply
appreciated. This work has been partially supported by the 2005
PATOM CNRS program and by 2005-ANR project TOPOGI-3D.

\end{document}